\documentclass[%
 amsmath,amssymb,
 aip,
 jcp,
 reprint,
 twocolumn,
]{revtex4-2}

\usepackage{graphicx}
\usepackage{dcolumn}
\usepackage{bm,color}
\usepackage{amsmath, amssymb, amsthm}
\usepackage{hyperref}
\usepackage[version=4]{mhchem}

\usepackage{xcolor}
\usepackage{soul}
\usepackage[most]{tcolorbox}
\tcbset{highlight math style={boxrule=0pt, colback=yellow, left=2pt, right=2pt, top=1pt, bottom=1pt}}
\begin{document}

\preprint{APS/123-QED}

\title{Field-driven Ion Pairing Dynamics in Concentrated Electrolytes}

\author{Seokjin Moon}
\affiliation{Department of Chemistry, University of California, Berkeley, CA, 94720,  USA}
\author{David T. Limmer}
\email{dlimmer@berkeley.edu}
\affiliation{Department of Chemistry, University of California, Berkeley, CA, 94720, USA}
\affiliation{Kavli Energy NanoScience Institute, Berkeley, CA, 94720, USA}
\affiliation{\mbox{Materials Sciences Division, Lawrence Berkeley National Laboratory, Berkeley, CA, 94720, USA}}
\affiliation{\mbox{Chemical Sciences Division, Lawrence Berkeley National Laboratory, Berkeley, CA, 94720, USA}}

\date{\today}

\begin{abstract}
We investigate ion pairing dynamics in electrolytes driven far from equilibrium using molecular simulations and nonequilibrium rate theory. Focusing on 0.5 M $\mathrm{LiPF_6}$ in water and acetonitrile under uniform electric fields, we compute transition path theory observables including reactive fluxes and mean first-passage times of ion pairing. Moreover, we introduce a dynamical proxy of free-ion population, where its field-induced change is strongly correlated with the nonlinear enhancement of conductivity, yielding an increase of $40 \ \%$ at $50 \ \mathrm{mV / \AA}$ in acetonitrile, compared to less than $10 \ \%$ in aqueous electrolytes. Further kinetic analysis elucidates that Onsager's classical theory substantially overestimates field-induced enhancement of ion pair dissociation in molecular electrolytes. This discrepancy arises from solvent-mediated dynamical pathways and field-induced dielectric decrement that suppress ion pair dissociation within explicit solvents, highlighting that a faithful description of molecular details is essential. Our results provide a molecular interpretation of nonlinear electrolyte transport beyond continuum theories and establish a general framework for quantifying nonequilibrium reaction kinetics in condensed phase systems.

\end{abstract}

\maketitle

\section{\label{sec:level1}Introduction}
Driven electrolytes have been at the
forefront of scientific and technological developments over the last century. Although significant theoretical work exists to describe their behavior under simplifying dilute solutions near equilibrium, much remains to be understood in molecular detail, at high concentration, and away from equilibrium.\cite{harned1950physical} Here, we study the ion pairing dynamics of concentrated electrolytes in a nonequilibrium steady state driven by strong electric fields. We develop and apply a rate theory to measure rate constants for diffusion-controlled ion pairing processes far from equilibrium. We introduce a dynamical free-ion population that explains the field-induced nonlinear increase in conductivity, namely the \emph{second Wien effect}.\cite{onsager1934deviations} Our molecular simulations reveal that ion dissociation rates increase with an applied field, and this effect is larger for less polar solvents whose dielectric decrement is smaller. Detailed dynamical analysis provides evidence that solvent degrees of freedom are essential for describing ion pairing kinetics.

The coupling between ion pairing and nonequilibrium transport has been studied since the beginning of modern physical chemistry, motivated by early observations of nonlinear conductivity in electrolytes subjected to strong electric fields.\cite{wien1927deviations, onsager1934deviations,marcus2006ion,kaiser2013onsager} In a seminal contribution,  Onsager provided a theoretical explanation of the second Wien effect by connecting the enhanced conductivity with a field enhanced concentration of free-ions.\cite{onsager1934deviations} Onsager derived an analytic expression for the field dependent dissociation constant of weak electrolytes as a universal function of two competing length scales set by inter-ion electrostatic interactions and the applied field.\cite{onsager1969motion} In modern contexts, these ideas continue to inform the interpretation of experiments in systems such as bipolar membranes for water electrolysis, where strong electric fields accelerate water dissociation and proton transport.\cite{parnamae2021bipolar,bui2024multi,chen2022design} Despite the increasing molecular complexity of these systems, experimental analysis often remains grounded in Onsager's original continuum-based expression.\cite{ortega2022modelling} For concentrated electrolytes, short-ranged interactions, explicit solvation, and correlated dynamics become increasingly important, and the assumptions underlying classical ion pairing theory can break down.\cite{roy2017marcus,limmer2015interfacial,dinpajooh2024beyond,coles2020correlation, kudlay2004complexation}


Modeling of ion pairing and transport in concentrated electrolytes under electric fields is most naturally pursued with molecular simulations. Yet even with nonequilibrium trajectories in hand, extracting quantitative kinetics remains nontrivial. For a system evolving far from equilibrium,  analysis frameworks tied to equilibrium free-energy calculations are not directly applicable.\cite{limmer2024statistical} Moreover, ion pairing events are often diffusive, and the separation between local environmental relaxation and the global association/dissociation process may not be strict, complicating application of standard rate theories that rely on a clear timescale separation.\cite{frenkel2023understanding,agmon1990theory} These challenges motive developments of theories that can extract kinetics directly from nonequilibrium trajectory ensembles, rather than relying on linear response theories.\cite{schile2019rate,kuznets2021dissipation,gao2017transport,gao2019nonlinear,limmer2021large}

\begin{figure*}[th]
\centering
\includegraphics[width=\textwidth]{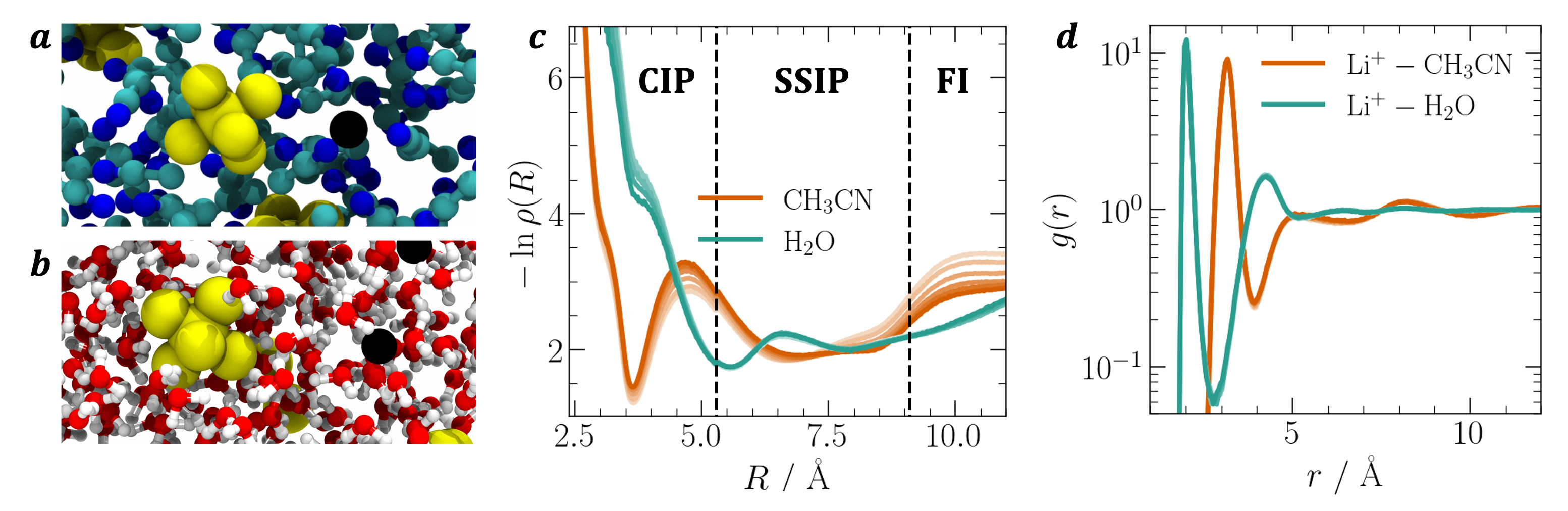}
\caption{\label{fig:1}Molecular dynamics simulation snapshots of $\mathrm{0.5 \ M}$ $\mathrm{LiPF_6}$ in (a) acetonitrile and (b) water. Black and yellow spheres represent $\mathrm{Li^+}$  and $\mathrm{PF_6^-}$, respectively. (c) Probability distribution function of the cluster-based nearest-counterion distance. Colors progress from light to dark with increasing field strength from $0$ to $50$ $\mathrm{mV / \AA}$. Dashed lines indicate locations of domain boundaries. (d) $\mathrm{Li^+}$-solvent  radial distribution functions with the same color gradient scale in c. }
\end{figure*}

We address these challenges by performing nonequilibrium molecular dynamics simulations of concentrated $\mathrm{LiPF_6}$ in water and acetonitrile under strong electric fields. These solutions serve as simple models of  relatively weak electrolytes in solvents of varying polarity. Building on transition path theory,\cite{vanden2006towards, berezhkovskii2022relations} we formulate a chemical rate theory within nonequilibrium steady states by defining committor-weighted densities that encode history dependence arising from finite solvent reorganization times.\cite{moon2025a} We then extract rate constants directly from the nonequilibrium trajectory ensemble via mean first-passage times, which remains well-defined without a clear separation of timescales. The trajectory-based analysis reveals how explicit solvation, dielectric response, and correlated dynamics reshape nonlinear transport, yielding a molecular interpretation of the second Wien effect.

\section{\label{sec:main1} Results and Discussion}
Representative snapshots of $\mathrm{0.5 \ M}$ $\mathrm{LiPF_6}$ simulations in two solvents are shown in Figs.~\ref{fig:1}a (acetonitrile) and ~\ref{fig:1}b (water). The systems are studied in cubic boxes with side lengths of $\mathrm{36.00 \ \AA}$ (acetonitrile) and $\mathrm{29.84 \ \AA}$ (water). Periodic boundary conditions are applied in all directions, yielding bulk electrolyte transport. A uniform electric field, $\xi$, is applied along the $z$ direction with strengths spanning $\mathrm{0 \ to \ 50 \ mV / \AA}$. The largest field is sufficient to induce significant dielectric decrement, while remaining below the regime associated with dielectric breakdown and autoionization for water.\cite{saitta2012ab} Simulations are carried out at fixed particle number $N$, volume $V$ and temperature $T = \mathrm{300 \ K}$ using a Langevin thermostat with a momentum relaxation time of $\mathrm{2 \ ps}$. Nonbonded interactions are described by Coulomb and Lennard-Jones potentials, with particle-particle particle-mesh long-range electrostatic solvers. Each simulation is relaxed for $\mathrm{10 \ ns}$ to reach a nonequilibrium steady state, followed by a $\mathrm{100 \ ns}$ production run. The SPC/fw model is used for modeling water,\cite{wu2006flexible} while all other species are described with OPLS-AA parameters.\cite{jorgensen1996development, muralidharan2018comparison} All simulations are performed using LAMMPS.\cite{plimpton1995fast, thompson2022lammps}

To begin, we seek to identify an order parameter that quantifies the population of free-ions in the electrolyte, which is expected to control the resultant DC-conductivity. In dilute electrolytes, the interionic separation often provides an adequate description. The potential of mean force along the interionic separation enables a distance-based classification of ion pairs into contact ion pairs (CIP), solvent-separated ion pairs (SSIP), and free-ions (FI),\cite{dang2017rate} which is consistent with assignments based on the $\mathrm{PF_6^-}$ vibrational signatures of ion pairing.\cite{fulfer2018ion}  In concentrated electrolytes, the interionic distance of a selected cation-anion-pair is poorly aligned with transport-relevant observables. For example, a two-body ion pair can collide with another free-ion to form a transient three-ion cluster. A definition based solely on interionic distance would classify this process as ion pairing, while not necessarily changing the population of charge-carrying free-ions that governs the  conductivity. This motivates a definition of reactive ion pairing events conditioned on changes in the free-ion population, explicitly accounting for clustering. Accordingly, we introduce a cluster-based nearest-counterion distance $R[\mathbf{r}^N]$, defined as the distance between a specifically chosen pair of $\mathrm{Li^+}$ and the central phosphorous on the anion, while $\mathbf{r}^N$ is the full configuration of the $N$ particle system. For an ion, we identify the nearest oppositely charged cluster and then select the nearest counterion within that cluster, such that changes in this distance corresponds to charge-reducing events classified as ion pairing events. Details on the order parameter are further discussed in Supplementary Methods S1.1. 

We measured the probability density $\rho(R)=\langle \delta(R-R[\mathbf{r}^N])\rangle_\xi$, where the brackets denote steady state average at field $\xi$. This is shown in Fig.~\ref{fig:1}c on a logarithmic scale. Both electrolytes exhibit clear structural features associated with molecular packing. We characterize a stable CIP for $R\leq R_\mathrm{i}=5.3 \ \mathrm{\AA}$, separated from a metastable SSIP with $R_\mathrm{i} \leq R \leq R_\mathrm{o}=9.1 \ \mathrm{\AA}$. Ions with $R>R_\mathrm{o}$ are thus free-ions. For the acetonitrile electrolyte, upon application of increasingly strong electric fields, the population of ions shifts systematically from paired states toward free-ion states. By contrast, for the aqueous electrolyte the CIP is only metastable due to the strong dielectric screening of water, which stabilizes dissociated ions and gives rise to a relatively stable SSIP. Applying an electric field to the aqueous system does not induce a noticeable population shift, except for a slight stabilization of the CIP.

\begin{figure}
\centering
\includegraphics[width=\columnwidth]{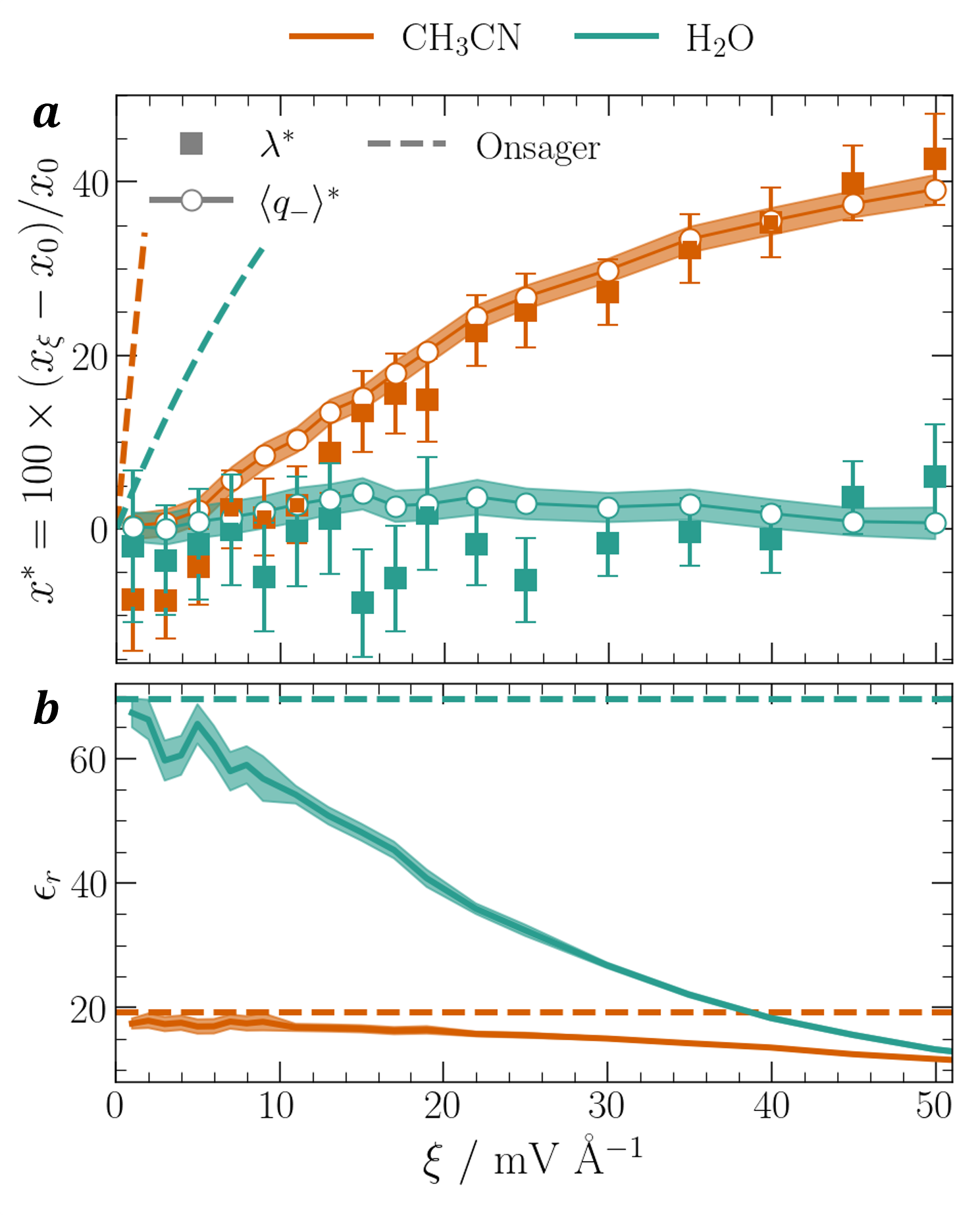}
\caption{\label{fig:2}(a) Relative percent change of the molar conductivity of electrolytes with applied field (filled symbols) and free-ion population (empty symbols). The values at zero field are $\lambda(0) = 66.5 \pm \ 1.9 \ \mathrm{S \ cm^2  \ mol^{-1}}$, $\lambda(0)= 78.4 \pm \ 2.6 \ \mathrm{S \ cm^2 \  mol^{-1}}$, $\langle q_- \rangle_0 = 0.401 \pm \ 0.002 $ and $\langle q_- \rangle_0 = 0.585 \pm \ 0.004 $ for acetonitrile (orange color) and water (blue color), respectively. (b) Relative dielectric constant in parallel direction to the electric field. Dashed lines indicate the zero field limits.}
\end{figure}

We characterize the solvation structure around Li$^+$ by computing its radial distribution function, $g(r)$,\cite{hansen2013theory} with the solvent, shown in Fig.~\ref{fig:1}d. For water, $r$ is defined as the distance between Li and the oxygen on water, while for acetonitrile it is between Li and the central carbon. For both solvents, the angle-averaged solvent coordination exhibits little sensitivity to the applied electric field. The average number of solvent molecules within the first solvation shell is $5.0$ for acetonitrile and $4.3$ for water and does not change appreciably over the field strengths we consider. We attribute this insensitivity of the solvation structure to the strong electrostatic interaction between the ion and the surrounding solvent dipoles, which dominate over the applied field.  

In the linear response regime, the electrolyte conductivity can be measured from equilibrium simulations via the fluctuation-dissipation relation.\cite{hansen2013theory} Here, we used the Einstein--Helfand equation for the zero-field limit of the molar conductivity,\cite{helfand1960transport}
\begin{equation}
    \lambda_0 = \frac{1}{6N_\mathrm{I} k_\mathrm{B} T}\lim_{t \rightarrow \infty} \frac{d\langle \left(\Delta \mathbf{ G}\right)^2 \rangle_0} {dt}, \quad \mathbf{G}=\sum_{i=1}^{2N_\mathrm{I}}q_i \mathbf{r}_i 
\end{equation}
where $\Delta \mathbf{G}(t)=\mathbf{G}(t)-\mathbf{G}(0)$ is the Helfand moment, $q_i$ is the charge of ion $i$ at position $\mathbf{r}_i$ and $N_\mathrm{I}$ is the number of cations or anions. Outside the linear response regime, we compute the molar conductivity from the current response to an applied field $\xi$,\cite{lesnicki2021molecular}
\begin{equation}
    \lambda_\xi = \frac{1}{N_I k_\mathrm{B} T}\frac{d\langle J_z \rangle_\xi }{d\xi}, \quad J_z = \frac{dG_z}{dt},
\end{equation}
where $J_z$ is the $z$-component of current vector. The field-dependent conductivities are shown in the Fig.~\ref{fig:2}a. Over the applied field range, we observed markedly different behaviors in the two solvents. The acetonitrile electrolyte exhibit a roughly $40 \ \%$ increase in conductivity at $\xi=50 \ \mathrm{mV / \AA}$, whereas the aqueous electrolyte shows an increase of less than $10\ \%$ at the same field. 

While our structure-based analysis provides a necessary first step toward a molecular understanding of speciation and nonlinear change in conductivity, it is not sufficient to describe the underlying ion pairing dynamics.\cite{balos2020macroscopic} A central challenge lies in assessing how conductivity correlates with different ion pair states. Although the CIP population remains largely unchanged under the applied field, we observe substantial changes in conductivity, indicating that the SSIP may play a central role in the nonlinear conductivity response. This observation motivates the introduction of dynamical descriptors, as static structural populations alone are insufficient to capture the role of SSIP in nonlinear conductivity.\cite{peters2016reaction} This limitation reflects the fleeting, metastable nature of SSIP and the importance of solvent reorganization beyond instantaneous interionic distances.

We collect nonequilibrium trajectories of the cluster-based nearest-counterion distance $R(\mathbf{r}^N)$ and use this coordinate to define two disjoint sets, $A$ and $B$, and an intermediate domain $\Omega_{AB}$.  Specifically, $B$, $\Omega_{AB}$, and $A$ correspond to CIP, SSIP, and free-ions, respectively, with classification criteria illustrated in Fig.~\ref{fig:1}c. 
%
With these definitions, we used the mean backward committor, $\langle q_- \rangle_\xi$, as a dynamical proxy of the free-ion population. The backward committor, $q_-(\mathbf{r}^N)$, is the probability that the assigned counterion in configuration $\mathbf{r}^N$ most recently visited the free-ion state rather than the CIP state. The mean is evaluated with respect to the nonequilibrium steady state distribution under applied field. This dynamic population includes all free-ions as well as a subset of SSIPs that originate from the free-ion state. By construction, $q_-$ incorporates trajectory history and thus captures the slow relaxation of the solvent following rapid ion pairing events. For example, typical lithium-acetonitrile exchange timescales are reported as $\sim 100 \ \mathrm{ps}$,\cite{dereka2022exchange} whereas the SSIP lifetime is estimated here to be $\sim 12 \ \mathrm{ps}$. The dynamic population and its field-dependence depend on the location of the $A$ and $B$ boundaries, and we have explored alternatives without qualitatively different observations. Details on our choice of parameters are discussed in Supplementary Notes S2.1. The field-dependent increase of the dynamic population shown in Fig.~\ref{fig:2}a strongly correlates with the observed conductivity enhancement. The pronounced increase of free-ions in acetonitrile thus provides molecular evidence for the second Wien effect in a concentrated electrolyte. The aqueous electrolyte exhibits a minor nonlinear response, and correspondingly little change in free-ion population. 

These observations are qualitatively consistent with Onsager's original analysis for dilute solutions. We consider the ion pairing reaction,
\begin{equation}
    \ce{Li+ + PF6- <=>[k_a][k_d] LiPF6}
\end{equation}
with $k_\mathrm{d}$ the dissociation rate constant and $k_\mathrm{a}$ the association rate constant. A dissociation constant, $K(\xi)$,
\begin{equation}
    K(\xi) = \frac{k_\mathrm{d}(\xi)}{k_\mathrm{a}(\xi)} = c_0 \frac{\langle q_- \rangle_\xi^2}{1-\langle q_- \rangle_\xi}
\end{equation}
relates the dynamic free-ion populations with $c_0=0.5 \ \mathrm{M}$ to the ratio of the pseudo-first-order rate constant $k_\mathrm{a}$  and the first-order rate constant $k_\mathrm{d}$. According to Onsager's analysis, the field-induced shift in the dissociation constant $K(\xi)$ has an analytical expression in terms of the ratio of the Bjerrum length $l_B=q^2/4\pi\epsilon k_\mathrm{B} T$ and the electrostatic length $l_E=k_\mathrm{B} T/q\xi$,\cite{onsager1934deviations} 
\begin{equation}
    \frac{K(\xi)}{K(0)} \approx \frac{I_1(\sqrt{4x})}{\sqrt{x}}, \label{eq:onsager}
\end{equation}
where $x=l_B/l_E$, $k_\mathrm{B}$ is Boltzmann's constant, $\epsilon$ is the dielectric constant, and $I_1$ is the modified Bessel function of the first kind. Therefore, the enhanced conductivity is estimated by field-induced increase of dynamic free-ion population implied by $K(\xi)$. 
As shown in Fig.~\ref{fig:2}a, this analysis does predict an increase in conductivity, but overestimates the nonlinear response for the concentrated electrolytes considered here. The classic successes of continuum theories, such as the application of Onsager's framework to weak acid dissociation\cite{eckstrom1939wien} and water autoionization\cite{cai2022wien}, may be rationalized by the chemical nature of these processes where covalent bonds are broken, which represent regimes that are now accessible to molecular simulations through advances in machine-learning force fields.\cite{joll2024machine}

\begin{figure}
\centering
\includegraphics[width=\columnwidth]{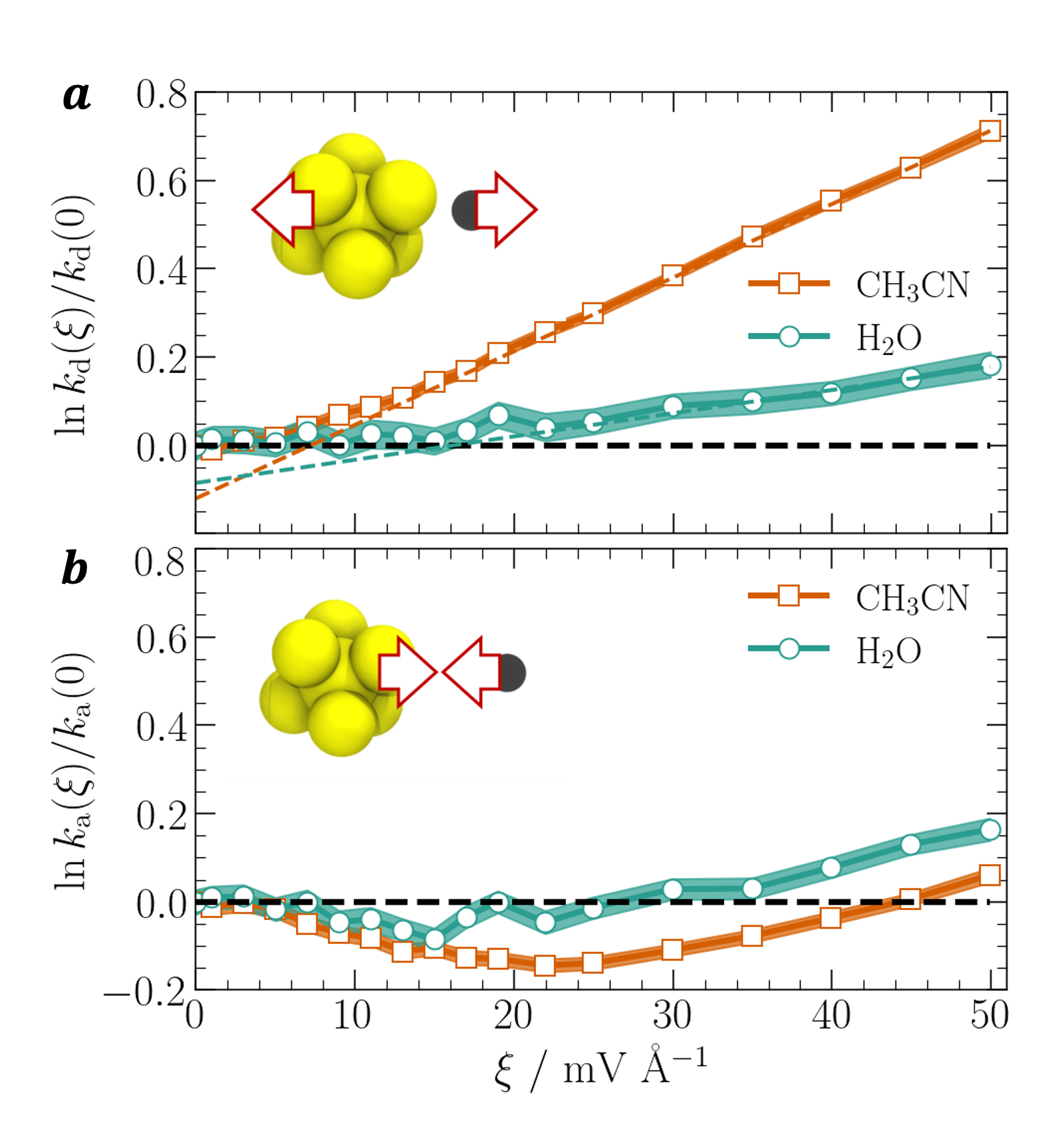}
\caption{\label{fig:3}(a) Field-induced dissociation and (b) association rate enhancements of the $\mathrm{LiPF_6}$ in acetonitrile (square) and water (circle) in log-scale. The rate constants at zero field for acetonitrile are
$k_{\mathrm{d}}(0) = 12.7 \ \pm \ 0.1 \ \mathrm{ns}^{-1}$ and $k_{\mathrm{a}}(0) = 18.9 \ \pm \ 0.1 \ \mathrm{ns}^{-1}$, and those for water are
$k_{\mathrm{d}}(0) = 36.2 \ \pm \ 0.4 \ \mathrm{ns}^{-1}$ and $k_{\mathrm{a}}(0) = 25.6 \ \pm \ 0.3 \ \mathrm{ns}^{-1}$. Colored dashed lines show a linear fit and black dashed line indicates zero field value.}
\end{figure}

Ion pairing is known to be determined by collective solvent motions that are significantly affected by the field while not captured by Onsager's continuum theory.\cite{geissler1999kinetic,mullen2014transmission,kattirtzi2017microscopic} In Fig.~\ref{fig:2}b, we show the parallel component of the differential dielectric constant along the field,
\begin{equation}
    \epsilon_r(\xi) = 1+ \frac{1}{\epsilon_0 V} \frac{d\langle M_z\rangle_\xi }{d\xi}, \quad M_z= \sum_{i=1}^{N-2N_\mathrm{I}} q_i z_i,
\end{equation}
where $M_z$ is the system dipole in the direction of the field. We also compute the value at zero field from a fluctuation relation, $\epsilon_r(0) = 1+  \langle \delta M_z^2\rangle_0 /\epsilon_0Vk_BT$. Both electrolytes show a clear dielectric decrement as a consequence of the field, with a more significant change for water than for acetonitrile. Such field-induced changes in dielectric response can weaken ionic solvation and thereby stabilize ion pairs, counteracting the dissociative driving force of the external field. To understand how dielectric decrement impacts the ion pairing dynamics, we require a kinetic description of association and dissociation.

To proceed, we employ a recently established connection between the reactive flux and the mean first-passage times, $\tau_{AB}(\mathbf{r}^N)$, for general nonequilibrium steady states.\cite{moon2025a} The reactive flux, $\nu_{AB}(\xi)$, is defined as the frequency of observing trajectories that go from $A$ to $B$ without returning to $A$ while traversing $\Omega_{AB}$. The connection is made by introducing dynamic populations,
\begin{equation}\label{eq:MFPT}
    \frac{\nu_{AB}(\xi)}{\langle q_- \rangle_\xi} = \langle \tau_{AB} \rangle_\xi^{-1} \equiv k_\mathrm{a}(\xi),
\end{equation}
which motivates identifying $\nu_{AB}(\xi)/\langle q_- \rangle_\xi$ as a generalization of the associate rate constant $k_\mathrm{a}(\xi)$. Analogously, the ratio of the reverse reactive flux $\nu_{BA}(\xi)$ to $1-\langle q_- \rangle_\xi$ is the generalized dissociation rate constant, $\nu_{BA}(\xi)/(1-\langle q_- \rangle_\xi)=k_\mathrm{d}(\xi)$. The steady state condition guarantees $\nu_{AB} = \nu_{BA}$, leading to a chemical balance equation valid for far from equilibrium systems.

The generalized kinetic framework enables measurement of ion pairing rates by directly counting reactive events. We note that the canonical rate measurement based on a single dividing surface can fail due to a lack of timescale separation. A traditional side-side correlation function\cite{chandler1978statistical} will exhibit non-linear behavior, leading to difficulty in extracting well-defined rates. The application of a single-boundary approach to our system is discussed in Supplementary Notes S2.2, supporting the necessity of transition path theory analysis. The resulting field-induced enhancements of the dissociation and association rate constants are shown in Fig.~\ref{fig:3}a and \ref{fig:3}b. In both electrolytes, the dissociation events become faster with stronger fields. However, the enhancement is substantially smaller in the aqueous electrolyte ($\sim 24 \ \%$) than in acetonitrile ($\sim 100 \ \%$) at $\xi = 50 \, \mathrm{mV / \AA}$. The dissociation rate increases monotonically with field, exhibiting a quadratic rise in the low-field regime followed by exponential growth at high fields. This behavior is both quantitatively and even qualitatively distinct from Onsager's prediction in Eq.~\ref{eq:onsager} which is linear near $\xi=0$, while resembling nonequilibrium barrier-crossing rate enhancements reported previously.\cite{kuznets2021dissipation, singh2025reactive} We interpret this observation separately in two regimes. At weak fields, the angular distribution of the CIP dipole is only weakly perturbed so the field enhances transition along the favorable direction while simultaneously suppressing oppositely directed transition, leading to quadratic enhancement. At stronger fields, dissociation becomes biased along the field direction, allowing the field to accelerate dissociation more effectively. However, the observed weak responses cannot be estimated from equilibrium-based field-induced free-energy-reduction picture of barrier-crossing events. 

The association rate constants exhibit a much weaker field dependence in both electrolytes compared to the dissociation rates. Therefore, although the systems studied here do not correspond to ideal weak electrolytes, the dominant contribution to the nonlinear transport arises from field-enhanced dissociation, consistent with Onsager's original physical argument.\cite{onsager1934deviations, onsager1969motion} Moreover, the field-induced change in association rates display non-monotonic behaviors. The rate initially decreases with increasing field and subsequently increases at higher field strengths. An analytical recombination theory for Coulomb-interacting ions predicts a monotonic increase of the association rate constant when the ion concentration is held fixed at infinite separation.\cite{onsager1938initial,isoda1994effect} By contrast, recent numerical simulations on implicit-solvent weak electrolytes report the same non-monotonic field dependence of the association rate observed here.\cite{moon2025a} The initial rate decrease in the association rate originates from depletion of ion pairs as the population shifts toward free-ions, whereas at high fields the rate increases again due to enhanced transport that accelerates ion encounters. This mechanism carriers over to the present concentrated electrolytes. However, the magnitude of the enhancement is much smaller than in the dilute implicit-solvent case, suggesting that explicit solvation suppresses this behavior\cite{moon2025a}.

\begin{figure}
\centering
\includegraphics[width=\columnwidth]{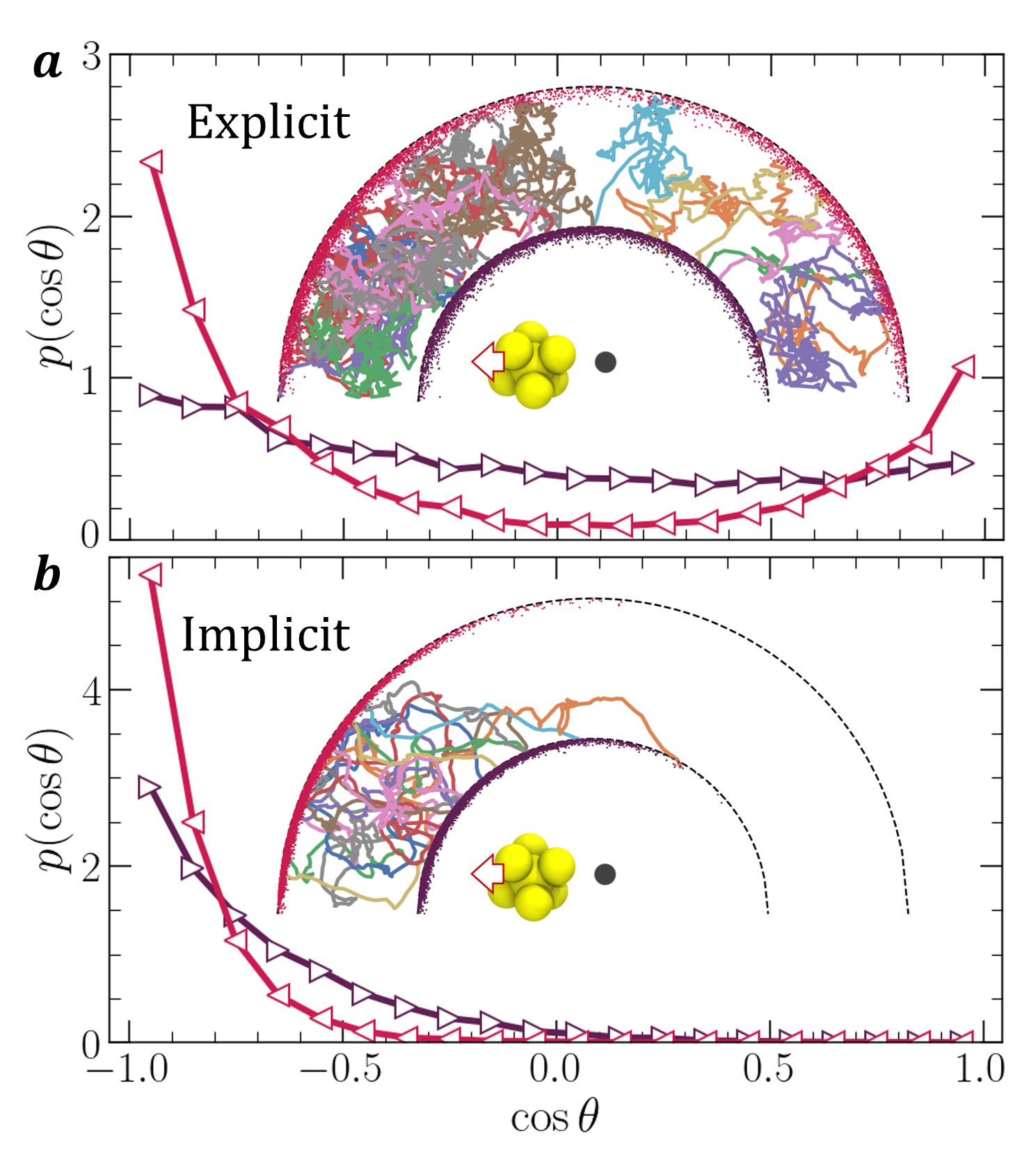}
\caption{\label{fig:4} The hitting probability distribution at boundaries (inner boundary for black and outer boundary for red) for the dissociation events for (a) explicit acetonitrile system and (b) implicit solvent system. The field strength is $50 \ \mathrm{mV / \AA}$ for both systems, pointing the right direction. Insets show representative reactive trajectories.}
\end{figure}

The hitting probability distributions of reactive trajectories along subdomain boundaries, or the angular histograms of counterion crossings, offer direct information about the most frequently visited reaction pathways. In Fig.~\ref{fig:4}a, we show the hitting probability distributions along the inner and outer boundaries for the dissociation process in the acetonitrile electrolyte, together with representative trajectory traces in the inset. For comparison, we also present corresponding results from implicit-solvent simulations with the same dielectric constant as acetonitrile in Fig.~\ref{fig:4}b. For both cases, $\cos\theta = \hat{z} \cdot \mathbf{R}/R$ where $\hat{z}$ is the unit vector along $z$ and $\mathbf{R}/R$ the unit vector along the inter-ion separation direction. The applied electric field biases the counterion motion toward the $\cos\theta = -1$ direction. 

In implicit-solvent, reactive pathways are strongly concentrated along the field, $\cos \theta =-1$, while trajectories crossing $\cos \theta =1$ are rare. This anisotropy becomes even more pronounced at the outer boundary, where the strong drift rapidly drives trajectories through the intermediate region. By contrast, explicit-solvent simulations exhibit a much more uniform angular distribution at the inner boundary. Although reactive trajectories remain biased along the field, crossing opposite of the field occur frequently. Upon reaching the outer boundary, trajectories continue to favor the lower-energy direction while retaining substantial weight along the higher-energy pathway, a feature absent in the continuum description. 

We interpret these differences as evidence that collective solvent fluctuations play an essential role in the dissociation process, rendering the interionic distance alone insufficient to describe molecular reactive pathways. In particular, solvent organization within SSIP aligns with the applied field, stabilizing the SSIP and giving rise to sluggish rotational dynamics. This finite rotational relaxation times of SSIP and even CIP in explicit-solvent naturally leads to a broader distribution of dissociation pathways, an effect not captured by continuum models.\cite{dalton2025memory} Dielectric relaxation spectroscopy measurements assigns a slow Debye mode to associated ions, reflecting the effective orientational relaxation of CIP or SSIP. For $\mathrm{1.0 \ M}$ $\mathrm{LiPF_6}$ in mixture of carbonate electrolytes with dielectric constant comparable to acetonitrile, reported relaxation times of of $\sim 300 \ \mathrm{ps}$ \cite{self2024solvation} are comparable to the estimated CIP lifetime here of $\sim 79 \ \mathrm{ps}$, supporting the importance of rotational dynamics in shaping dissociation pathways.

\section*{Conclusions}

In this work, we have shown that nonlinear transport in electrolytes driven far from equilibrium is governed by molecular ion pairing mechanisms that cannot be captured from continuum theories. Using molecular simulations combined with transition path theory, we demonstrated that a field-induced increase in the dynamical population of free-ions underlies the enhancement of conductivity, providing a molecular proof of the second Wien effect. We find that the dominant contribution to nonlinear transport arises from field-enhanced ion dissociation rate, while field-induced changes in the association rate play a secondary role. Our results further show that explicit solvent dynamics, through dielectric decrement and solvent-mediated reaction pathways, substantially suppress dissociation rate enhancement and lead to weaker nonlinear responses, especially in aqueous electrolytes. These solvent-controlled mechanisms explain both the quantitative and qualitative deviations from Onsager's continuum-level predictions. More broadly, our findings highlight the need to incorporate molecular-level solvent dynamics into theories of nonequilibrium transport. 

Beyond electrolytes, the framework developed here provides a general approach for extracting reaction kinetics from nonequilibrium steady states in condensed phase systems. By defining committor-based dynamical populations rather than static populations, this approach reduces sensitivity to the specific choice of reaction coordinate and naturally accounts for diffusive dynamics characterized by frequent recrossing and slow environmental relaxation. In nonequilibrium steady states, sampling rare reactive events remains challenging in general and developing systematic strategies for this regime is an important direction for future work.\cite{singh2025variational,dickson2010enhanced} However, for diffusion-controlled processes where reactive events occur on comparatively short timescales, molecular simulations yield sufficient sampling to reliably extract kinetic observables, making the present framework directly applicable. Representative examples include surface adsorption and desorption,\cite{riechers2016surfactant} chemical reactions at interfaces,\cite{wang2025selective} and ligand-receptor binding on cell membranes.\cite{chen2001selectin}

\section*{ACKNOWLEDGMENT}
This work was supported by the Condensed Phase
and Interfacial Molecular Science Program (CPIMS) of
the U.S. Department of Energy under contract no. DEAC02-05CH11231. The authors are grateful to Leonardo Coello Escalante for useful discussions.

\section*{References}
\bibliographystyle{aipnum4-2}
\bibliography{reference}

\end{document}


\title{Supplementary Information for:\\ Field-driven Ion Pairing Dynamics in Concentrated Electrolytes}

\author{Seokjin Moon}
\affiliation{Department of Chemistry, University of California, Berkeley, CA, 94720,  USA}
\author{David T. Limmer}
\email{dlimmer@berkeley.edu}
\affiliation{Department of Chemistry, University of California, Berkeley, CA, 94720, USA}
\affiliation{Kavli Energy NanoScience Institute, Berkeley, CA, 94720, USA}
\affiliation{\mbox{Materials Sciences Division, Lawrence Berkeley National Laboratory, Berkeley, CA, 94720, USA}}
\affiliation{\mbox{Chemical Sciences Division, Lawrence Berkeley National Laboratory, Berkeley, CA, 94720, USA}}

\date{\today}

\maketitle

\newpage

\section{Supplementary Methods}
\subsection{Cluster-based Nearest-Counterion Distance Algorithm}
The application of transition path theory to concentrated electrolytes requires a partitioning of the full configurational state space $\mathbf{r}^N$ into physically meaningful subdomains that reflect proximity-based ion-pairing motifs, such as contact ion pairs, solvent-separated ion pairs, and free ions. As discussed in the main text, order parameters based solely on nearest-neighbor distances provide a convenient geometric measure of ionic proximity, while reactive events defined through such order parameter are not consistently correlated with changes in the population of charge carriers. This limitation becomes pronounced in concentrated regimes where ions frequently participate in transient ion clusters. These considerations motivate the introduction of an alternative order parameter that explicitly accounts for cluster connectivity and its impact on the effective charge carrier density.

There is no unique choice of an order parameter that incorporates cluster effects in concentrated electrolytes. Here, we introduce a simple and physically motivated construction, which we term the cluster-based nearest-counterion (CBNC) distance. The central idea is to define, for each ion, a nearest counterion that is selected based on both spatial proximity and cluster connectivity. At each time frame, ions are grouped into clusters based on a distance-based connectivity criterion with a cutoff of $\mathrm{5.3 \ \AA}$. Within each cluster, ions are assigned a role according to the net cluster charge. Specifically, for a cluster with total charge $+q$, $q$ cations are designated as \emph{singleton}, representing excess charge carriers, while the remaining ions in the cluster are designated as \emph{paired} ions. For ions designated as \emph{paired}, the CBNC distance is defined as the nearest oppositely charged ion within the same cluster, reflecting local neutralization. For \emph{singleton} ions, the CBNC distance is defined with respect to the nearest oppositely charged cluster, and the nearest counterion within that cluster. This construction ensures that the CBNC distance tracks the proximity of physically relevant counterions while remaining sensitive to changes in the effective charge carrier population.

For a system with $N_I$ cations and $N_I$ anions, CBNC construction yields $N_I$ distinct trajectories for each ionic species, each tracking the time-dependent proximity between a tagged ion and its assigned counterion partner. Because the identity of the assigned counterion may change as cluster connectivity evolves, naive implementation of the CBNC can exhibit apparent discontinuities associated with partner switching. In particular, such discontinuities may manifest as abrupt transitions between free-ion and contact ion pair states that bypass the solvent-separated ion pair region. To prevent these artificial jumps, we incorporate a relabeling rule directly into the definition of CBNC distance. When simultaneous associative/dissociative jumps occur, particle indices are reassigned so as to preserve continuity of the CBNC distance. This procedure enforces a sequential transitions through solvent-separated configurations, ensuring that the subdomains used in the TPT analysis remain disjoint, as required by the theoretical framework.

\section{Supplementary Notes}
\subsection{Choice of Transition Path Boundaries in Diffusive Ion Pairing Systems}
Transition path theory requires the definition of two disjoint subsets of the full state space that establish the initial and final states of a reactive process. In this work, these boundaries are specified using the cluster-based nearest-counterion distance as the order parameter. In concentrated electrolytes, the relevant state space is high-dimensional, and the resulting subdomains are generally non-spherical and strongly influenced by collective ionic and solvent correlations. 

In diffusive systems, the precise placement of boundaries is particularly important for quantitative analysis. While boundary definitions have only a weak effect on observables in high barrier crossing transitions due to the low probability of configurations near the barrier top, they can substantially influence kinetic observables when the barrier is shallow. Boundary choices that are physically motivated by the classification of contact ion pairs, solvent-separated ion pairs, and free ions are sufficient to qualitatively capture the strong correlation between field-induced enhancements of conductivity and the mean backward committor. The remaining question is how to choose boundary locations that render this relationship quantitative, and whether such choices align with physical intuition.

\begin{figure}
\centering
\includegraphics[width=4 in]{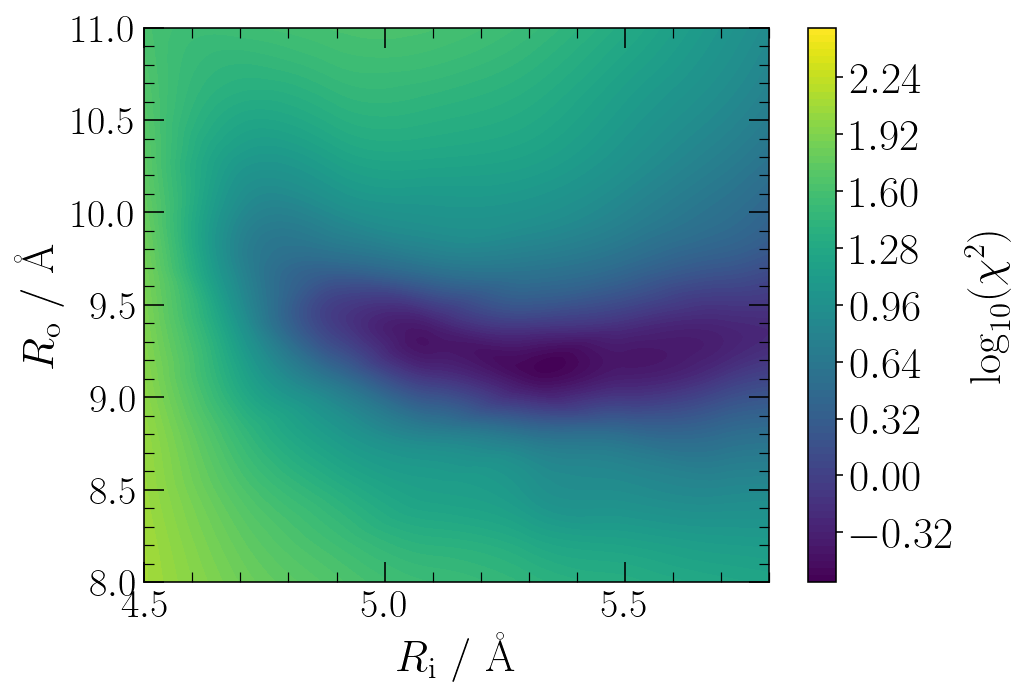}
\caption{\label{fig:s1} Deviation between conductivity ratios and boundary-dependent dynamic free ion population ratios. Darker regions indicate smaller deviations and thus more optimal choice of the boundaries.}
\end{figure}

To address this, we performed additional simulations of $\mathrm{LiPF_6}$ in acetonitrile with a slightly modified Lennard--Jones interaction between ions ($\sigma_{\mathrm{Li-P}} = 4.505 \ \mathrm{\AA},$ $\sigma_{\mathrm{Li-F}} = 4.113 \ \mathrm{\AA}$), resulting in a reduced population of contact ion pairs. For this modified system, we computed the molar conductivity at zero field and at $\xi=40 \ \mathrm{mV/\AA}$, together with the dynamic free-ion population. We then constructed three ratios of conductivities and three corresponding ratios of dynamic populations. One is the field enhancement in the original system. Another is the field enhancement in the modified system. The other is the ratio between the two systems at zero field. We introduced an error-weighted deviation metric $\chi^2$ that quantifies the discrepancy between conductivity ratios and dynamic free-ion population ratios,
\begin{equation}
    \chi^2 = \sum_{j=1}^{3} \left( \frac{f_j(R_\mathrm{i}, R_\mathrm{o} ) - y_j }{\sigma_{y_j}} \right)^2, 
\end{equation}
where $f_j$ are three ratios of dynamic free-ion populations that depend on the boundary locations, $y_j$ are three ratios of conductivities, and $\sigma_{y_j}$ are the standard errors of conductivity measurements. By scanning over possible boundary locations, we identified the boundary pair that minimizes this metric. (Supplementary Fig.~\ref{fig:s1}) The dominant sensitivity arises from the placement of the outer boundary, which is optimally located at $R_\mathrm{o} = 9.1 \ \mathrm{\AA}$. In contrast, the inner boundary exhibits a degree of degeneracy around $R_\mathrm{i} = 5.3 \ \mathrm{\AA}$, reflecting its weaker influence on the observables. This asymmetry arises because the outer boundary more strongly affects population measures through the associated Jacobian factor in configurational space.

Importantly, the optimal boundary locations coincide with physically meaningful separations suggested by the probability distribution of the order parameter in Fig.~1c. This consistency supports the use of these boundaries in the main analysis and indicates that the chosen boundary conditions provide a quantitatively robust and physically grounded description of diffusive ion pairing dynamics.

\subsection{Limitations of Single-boundary Rate Measurements for Diffusive Ion Pairing}
A common approach for investigating equilibrium ion pairing in dilute limit is to apply standard free energy calculations in combination with rate theory. Such approaches rely on a clear separation of timescales between the global transition event and the local relaxation of the surrounding environment. This timescale separation guarantees the existence of a linear-response regime in the side-side correlation function $C_{AB}(t)$, corresponding to the exponential decay of density fluctuation following an external perturbation.
\begin{equation}
    C_{AB}(t) = \frac{\langle h_A[\mathbf{r}^N(0)] h_B [\mathbf{r}^N(t)] \rangle }{\langle h_A \rangle }, \quad h_A = 1-h_B =
    \begin{cases}
    1, & \mathbf{r}^N(t) \in A, \\
    0, & \text{otherwise}.
\end{cases}
\end{equation}

\begin{figure}
\centering
\includegraphics[width=6.5 in]{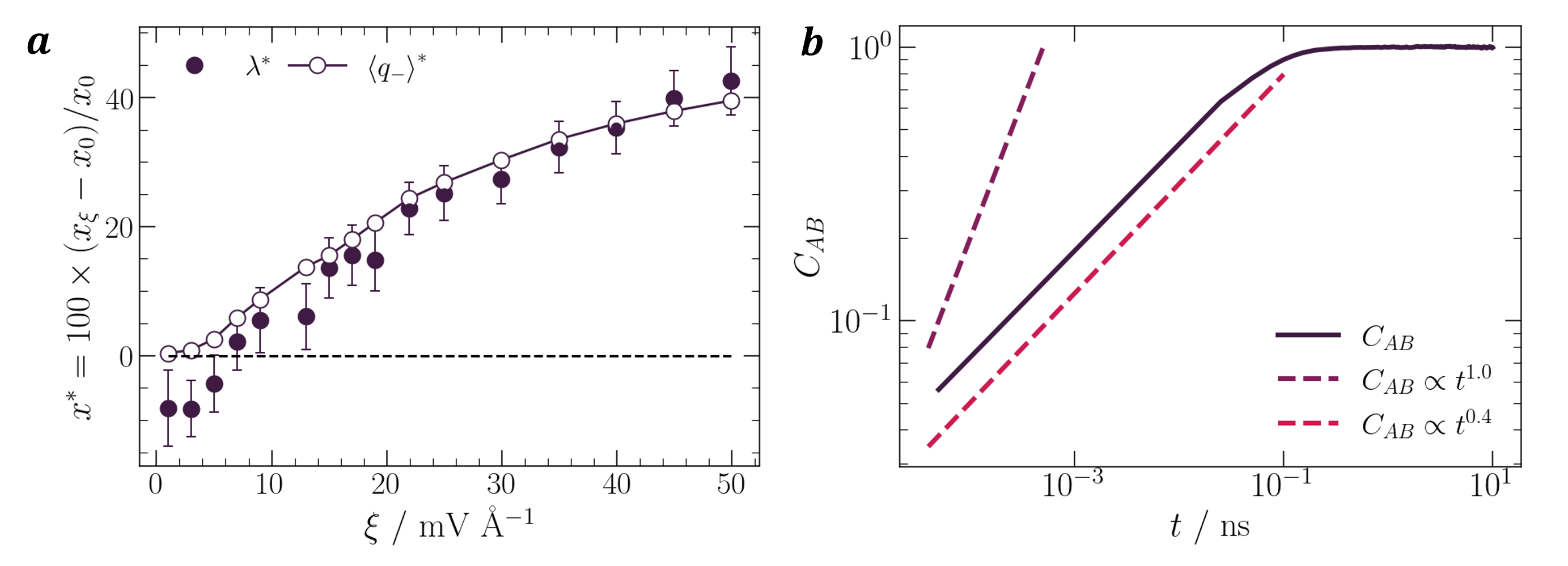}
\caption{\label{fig:s2} (a) Correlation between field-induced conductivity enhancement and free-ion population enhancement for  $\mathrm{LiPF_6}$ in acetonitrile when a single boundary at $R=7.2 \ \mathrm{\AA}$ is used. (b) Side-side correlation function plotted on a logarithmic scale. The relaxation exhibits sub-linear time dependence, approximately scaling as $t^{0.4}$, indicating absence of a well-defined linear response regime.}
\end{figure}

Under these conditions, the rate constant can be computed in an elegant and systematic manner by separating the free energy contribution from the dynamical correction, using the Bennett--Chandler formalism. However, in diffusive systems this framework encounters fundamental limitations. The absence of a well-defined timescale separation makes it difficult to identify a linear-response regime in the side-side correlation function. Instead, diffusive dynamics generally produce non-exponential relaxation behavior, which limits the extraction of a meaningful rate constant.

This breakdown is clearly observed in our simulations as shown in Supplementary Figs.~\ref{fig:s2}. Although it is possible to identify a single boundary location on the cluster-based nearest-counterion distance that reproduces the field-induced enhancement of conductivity as seen in Fig.~\ref{fig:s2}a, the corresponding side-side correlation function shown in Fig.~\ref{fig:s2}b exhibits pronounced non-linear behavior. This inconsistency highlights the inadequacy of single-boundary descriptions for diffusive ion-pairing dynamics and emphasizes the necessity of introducing two boundaries within the transition path theory framework.